\documentclass{mem}
\usepackage{natbib}\usepackage{txfonts}\usepackage{balance}
\usepackage{flushend}
\usepackage{graphicx}
\usepackage{comment}
\usepackage[breaklinks,pdftex]{hyperref}
\idline{94}{247}
\begin{document}
\def\teff{$T\rm_{eff }$}
\def\kms{$\mathrm {km s}^{-1}$}

\title{
Higher order clustering of Ly$
\alpha$ forest
}

   \subtitle{}

\author{
Soumak \,Maitra\inst{1} 
          }

\institute{
Istituto Nazionale di Astrofisica --
Osservatorio Astronomico di Trieste, Via Tiepolo 11,
I-34131 Trieste, Italy.
\email{soumak.maitra@inaf.it}
}

\authorrunning{Maitra }

\titlerunning{3-point correlation of Ly$
\alpha$F}

\date{Received: 29-10-2022; Accepted: 11-12-2022}

\abstract{
Higher order clustering statistics of Ly$\alpha$ forest provide a unique probe to study non-gaussianity in Intergalactic matter distribution up to high redshifts and from large to small scales. The author presents a brief review of his work studying the spatial clustering properties of Ly$\alpha$ absorbers, with emphasis on 3-point statistics. The observational side of this involves redshift-space clustering of low-$z$ ($z<0.48$) and high-$z$ ($1.7<z<3.5$) Ly$\alpha$ absorbers. This is complemented with astrophysical inferences drawn from N-body hydrodynamical simulations. We also use simulations to study 2-point and 3-point clustering statistics in the transverse direction using projected QSO triplet sightlines. Such studies will become possible observationally with upcoming surveys.
}
\maketitle{}

\section{Introduction}
The intergalactic gas comprising of the majority of the baryonic content of the Universe, is a known tracer of dark matter density fluctuations at large scales. At small scales, its matter distribution is influenced by gas pressure and other astrophysical processes. Ly$\alpha$ forest absorption in the sightlines of distant luminous quasar spectra provides a unique probe into this intergalactic matter distribution over a wide redshift range (from $z\sim 0$ using space based telescopes to $z\sim 6$). Clustering studies of these Ly$\alpha$ absorbers have played crucial role in improving cosmological and astrophysical understanding of the Universe. Its 2-point correlation function (2PCF) has been explored widely in redshift space along the quasar sightlines as well as in transverse direction using projected quasar pairs. Higher order clustering, however, remains largely unexplored in Ly$\alpha$ forest. Here, I will present a brief overview of my work studying the first higher order term in clustering statistics, i.e., 3-point correlation function (3PCF) of Ly$\alpha$ absorbers.

\begin{figure*}[t!]
\includegraphics[viewport=5 24 735 180,width=13.5cm,height=2.9cm, clip=true]{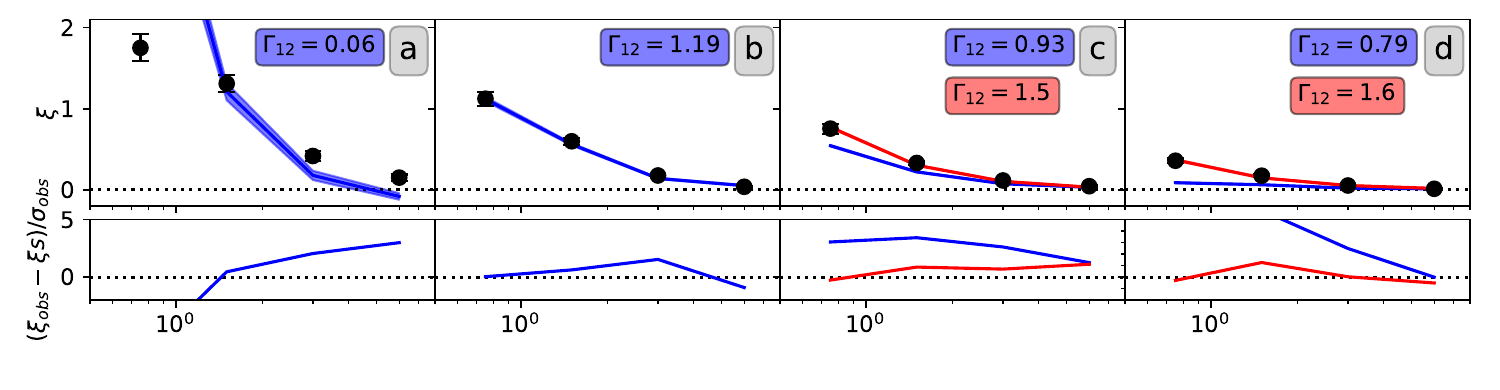}
    \includegraphics[viewport=5 5 735 183,width=13.5cm,height=3.1cm, clip=true]{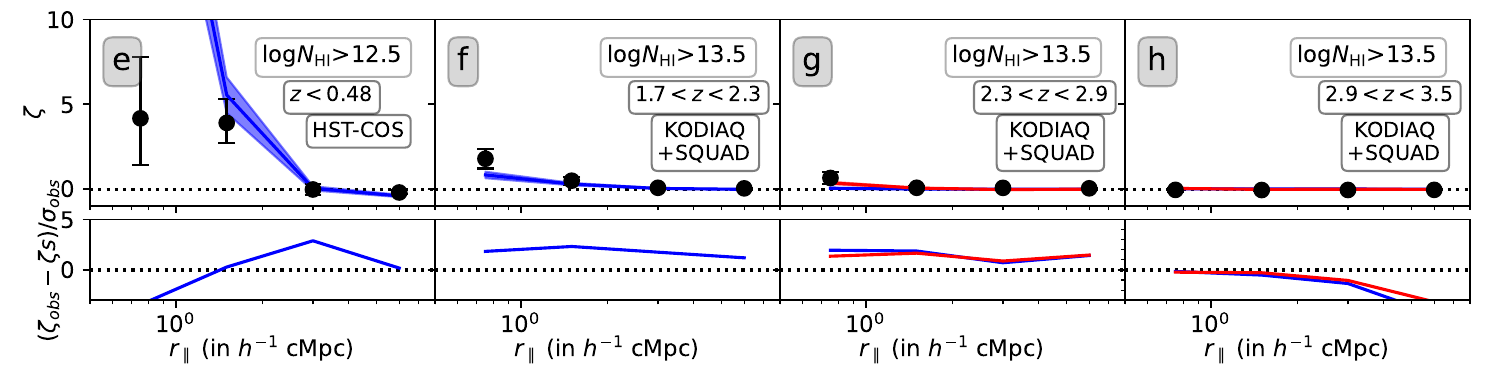}
\caption{\footnotesize Redshift-space 2PCF (\textit{Panel a-d}) and 3PCF (\textit{Panel e-f}) profile of Ly$\alpha$ absorbers. Black points show observed correlation functions obtained from {\sc HST-COS} ($z<0.48$) and joint {\sc KODIAQ+SQUAD} sample ($1.7<z<3.5$). Blue curves show correlation function obtained from {\sc GADGET-3} simulation (MassiveBlack-II (MBII) simulation \citep{khandai2015} used in the low-$z$ regime) using $\Gamma_{\rm HI}$ from \citet{khaire2019}. The red curve represent correlation function with elevated $\Gamma_{\rm HI}$ ($\Gamma_{12}$ values given). Additionally, plots of the difference in correlation profile between observed (obs) and simulated (s; color-coded) spectra relative to the error has also been plotted in the bottom.
}
\label{eta}
\end{figure*}

\section{Correlation estimator}

Clustering in Ly$\alpha$ forest has been widely estimated using pixel-based statistics, correlating transmitted flux ($F$) deviation field $F-\overline{F}$. This approach is largely affected by complex non-linear relation between $F$ field and matter density field. In the case of 3PCF in $F$, the amplitude is largely dominated by this complex relation rather than the non-gaussianity in matter distribution itself \citep{viel2004a}. Alternatively, thanks to the availability of high-resolution spectra, we use an absorber-based analysis for our clustering studies in \cite{maitra2019, maitra2020,maitra2020b,maitra2021}. Ly$\alpha$ absorption decomposed into Voigt profile components (using {\sc VIPER}; \cite{gaikwad2017b}) provide a straightforward way for estimating 3PCF and studying it as a function of HI column density ($N_{\rm HI}$) and line-width parameter ($b$). Their 2PCF ($\xi$) and 3PCF ($\zeta$) are then expressed using natural estimators \cite{kerscher2000}
\vspace{-0.4em}
\begin{equation}
		\xi( r_{\parallel})=\frac{\overline{DD}}{\overline{RR}}-1 \ ,
		\label{Corr2_eqn}
\end{equation}
\vspace{-0.5em}
\begin{equation}
    		\zeta( r_{\parallel})=\frac{\overline{DDD}}{\overline{RRR}}-1-\xi(r_1)-\xi(r_2)-\xi(r_3) \ ,
		\label{Corr3_eqn}	
	\end{equation}
	
where ``DD", ``RR" are data-data and random-random pair counts, and ``DDD", ``RRR" are data-data-data and random-random-random triplet counts. This estimator choice over more robust ones is justified due to reduced computation times. Also, clustering amplitudes remain relatively independent of estimator choice at small scales.

\section{Redshift space clustering}
\cite{maitra2020b} studies redshift space clustering of absorbers in $z<0.48$ Ly$\alpha$ forest using 82 high-resolution quasar spectra from HST-COS \citep{danforth2016}. We extended this study to $1.7<z<3.5$ Ly$\alpha$ absorbers in \cite{maitra2021} using 292 high-resolution spectra from {\sc KODIAQ} (KECK/HIRES) and SQUAD (VLT/UVES) survey. Fig.~\ref{eta} shows the consolidated clustering results.

\subsection{Low-$z$ ($z<0.48$) clustering}
\cite{maitra2020b} reports the first detection of non-zero 3PCF  ($\zeta =4.76^{+1.98}_{-1.67}$) for $N_{\rm HI}>10^{12.5}$cm$^{-2}$ Ly$\alpha$ absorbers at a scale of 1-2 pMpc (physical) with reduced 3-point correlation Q = $0.95^{+0.39}_{-0.38}$.  The amplitude of $\zeta$ increases with $N_{\rm HI}$ while Q remains relatively unaffected. {Frequency of occurrence of high-$b$
absorbers ($b>40km/s$, corresponding to $T>10^5K$ WHIM regions) is a factor $\sim3$ higher ($\sim$85\%) in triplets than that found among the full sample ($\sim$32\%), suggesting that Ly$\alpha$ triplets may preferentially be probing large scale filamentary structures. We also found that $\sim$88\% of the triplets contributing to $\zeta$ {(at $z\le 0.2$)} have nearby galaxies. The measured impact parameters are consistent with appreciable number of triplets not originating from individual galaxies but tracing the underlying galaxy distribution.} Simulation [MassiveBlack-II (MBII) \citep{khandai2015}] suggests that both $\xi$ and $\zeta$  are amplified strongly by redshift space distortion effects, implying presence of convergent flows. However, wind and AGN feedbacks have minor effects on the observed clustering.

\subsection{Intermediate-$z$ ($1.7<z<3.5$) clustering}

We measure positive 2-point correlation for $N_{\rm HI}>10^{13.5}$cm$^{-2}$ absorbers up to 8 $h^{-1}$ cMpc and study its evolution in the redshift bins $1.7<z<2.3$, $2.3<z<2.9$ and $2.9<z<3.5$. We detect positive $\zeta$ in $1.7<z<2.3$ and $2.3<z<2.9$ bins only upto  2$h^{-1}$cMpc  and 1$h^{-1}$cMpc, respectively. The strongest detection of $\zeta$ is seen in $z=1.7- 2.3$ redshift bin at $1-2h^{-1}$ cMpc with an amplitude of $1.81\pm 0.59$ ( $\sim 3.1\sigma$ level). 
The corresponding Q is found to be $0.68\pm 0.23$. Both $\xi$ and $\zeta$ show strong redshift evolution. We find that the strong redshift evolution shown by $\xi$ and $\zeta$ is primarily sourced by the redshift evolution of the relationship between $N_{\rm HI}$ and baryon overdensity ($\Delta$).
We also find that high-$z$ $\xi$ and $\zeta$ show an increasing trend with $N_{\rm HI}$ while Q remains almost unaffected. Simulations show that peculiar velocities amplify the correlation amplitudes. Simulations also show that feedback, thermal and pressure smoothing effects influence clustering of Ly$\alpha$ absorbers at small scales ($<0.5h^{-1}$cMpc). HI photo-ionization rates ($\Gamma_{\rm HI}$) strongly influence correlation amplitudes at all scales. Using existing fits for $\Gamma_{\rm HI}(z)$ measurements \citep{khaire2019}, the simulations produce consistent clustering signals with observations at $z\sim 2$ but underpredicts clustering at higher $z$. One possible solution is to have a slightly elevated $\Gamma_{\rm HI}$ at higher $z$. The issue could also be related to presence of non-equilibrium physics and inhomogeneity during HeII reionization, which our simulations do not capture.

\begin{figure}
\includegraphics[viewport=5 10 430 215,width=6.9cm,height=3.5cm, clip=true]{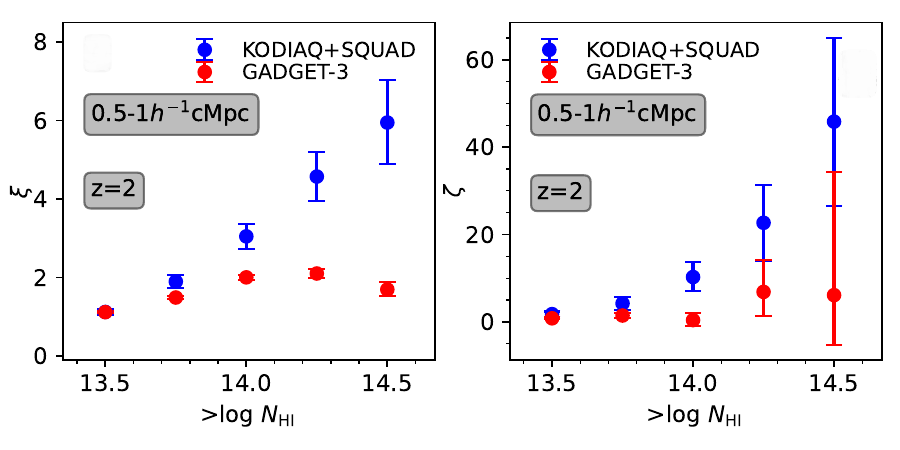}
\caption{\footnotesize Dependence of redshift-space 2PCF on $N_{\rm HI}$ thresholds for scales $0.5-1h^{-1}$cMpc. Blue points denote observational 2PCF corresponding to KODIAQ+SQUAD survey in $z=1.7-2.3$ while red points correspond to GADGET-3 simulation at z=2.
}
\label{NHI}
\end{figure}

\subsection{Comparison with simulations}
In the low-$z$ regime, the simulations reproduce the observed 2PCF and 3PCF profiles reasonably well at scales $>1h^{-1}$cMpc (simulated 2PCF profile being slightly steeper (see Fig.~\ref{eta})). At smaller scales, the simulations do not show the characteristic suppression seen in the clustering signals of absorbers. This suppression is associated with the difficulty in detection of multiple components within certain length scales due to finite width of absorbers. The disagreement between observation and simulation can likely be attributed to well-known issue of simulations producing absorbers with lower $b$ values (\cite{viel2016,gaikwad2017b,Nasir2017}).

In the intermediate-$z$ regime, the observed clustering signals are well reproduced in the simulations around $z\sim 2$ using KS19 (\cite{khaire2019}) UV background model, but require elevated HI photo-ionization rates for higher redshifts. This might suggest a need to revisit the existing $\Gamma_{\rm HI}$ measurements with an exhaustive parameter estimation process involving simultaneously matching various other observables including the absorber statistics.

Another issue involving the clustering of Ly-$\alpha$ absorbers is found towards the higher $N_{\rm HI}$ end. While, the simulations perform reasonably well in reproducing the observed clustering statistics with lower $N_{\rm HI}$ thresholds ($N_{\rm HI}>10^{12.5}$cm$^{-2}$ for low-$z$, $N_{\rm HI}>10^{13.5}$cm$^{-2}$ for intermediate-$z$), it generally tends to underproduce the observed clustering signals for higher thresholds. This behaviour has been shown in Fig.~\ref{NHI} at $z=2$ for $0.5-1h^{-1}$cMpc.

\section{Transverse clustering}
\cite{peeples2010a,peeples2010b} had pointed out that transverse correlation is more sensitive to pressure broadened 3D structures of IGM in comparison to redshift-space correlation (which is more sensitive to thermal broadening).
Hence, we used GADGET-3 simulation to generate 4000 mock spectra for  each projected quasar triplet configuration and analyze the 3PCF over a transverse scale of 1-5$h^{-1}$cMpc at $z\sim 2$.
We found that transverse 3PCF depends strongly on $N_{\rm HI}$ and scale and weakly on angle of the triplet configuration. We also obtain a median Q in the range 0.2–0.7 (mean in the range 0.4–1.3). We show the influence of physical parameters on Q is much weaker than that on $\zeta$.
Using simulations, we found transverse 2PCF and 3PCF to be a better probe of the thermal history (stronger effect of pressure broadening) in comparison to redshift space correlations.

\section{Conclusions}
A brief overview of the author's work in 3-point clustering of Ly$\alpha$ forest is presented. Using an absorber-based approach, we report the first detections of 3-point correlation in $z<0.48$ and $1.7<z<3.5$ Ly$\alpha$ forest at scales $\sim 1h^{-1}$cMpc. In the low-$z$ regime, we find indications for Ly$\alpha$ triplets to be originating preferentially from large scale filamentary structures in comparison to individual absorbers. In the intermediate-$z$ regime, simulations suggest the clustering to be primarily affected by background photo-ionization rates rather than thermal parameters. While the simulations more or less reproduce the 2PCF and 3PCF of Ly$\alpha$ absorbers in general, there are still some issues which needs to be addressed: (1) mismatch in correlation profiles at scales $<1h^{-1}$cMpc in low-$z$ regime, possibly due to simulations producing absorbers with lower $b$ values, (2) lower clustering signals for high $N_{\rm HI}$ absorbers and (3) lower clustering signals for $z>2.3$ Ly$\alpha$ forest.
The simulations have been helpful in analyzing 3-point statistics of Ly$\alpha$ forest in the transverse direction using projected quasar triplet sightlines which would be more helpful in probing the actual 3D structures of the IGM and its thermal history. While currently limited observationally, studying 3PCF in transverse direction is expected to become possible with upcoming surveys like DESI, LSST, etc.

\begin{acknowledgements}
The author acknowledges support from PRIN INAF - NewIGM project titled “Measurement of cosmological and astrophysical parameters from
the comparison of high resolution spectra of the intergalactic medium with simulations”. The author also acknowledges the use of HPC facilities {\sc perseus} and {\sc pegasus} at IUCAA-Pune for his work. The author also extends his gratitude to his Ph.D. supervisor Raghunathan Srianand and several others for their help and fruitful discussions.

\end{acknowledgements}
\bibliographystyle{aa}
\bibliography{Bibliography}

\end{document}